\documentclass[epj]{svjour}

\usepackage{graphicx,latexsym,xcolor}

\usepackage{amsmath}

\title{Entanglement and edge effects in superpositions of many-body Fock states with spatial constraints}

\author{Ioannis Kleftogiannis\inst{1} \and Ilias Amanatidis\inst{2} }

\institute{                    
  \inst{1} Physics Division, National Center for Theoretical Sciences, Hsinchu 30013, Taiwan\\
  \inst{2} Department of Physics, Ben-Gurion University of the Negev, Beer-Sheva 84105, Israel
}

\abstract{
We investigate how entangled states can be created by considering collections of point-particles arranged at different spatial configurations, i.e., Fock states with spatial constraints. This type of states can be realized in Hubbard chains of spinless hard-core bosons, at different fillings, which have gapped energy spectrum with a highly degenerate ground state. We calculate the bipartite entanglement entropy for superpositions of such Fock states and show that their entanglement can be controlled via the spatial freedom of the particles, determined by the system filling. In addition we study the effect of confinement/boundary conditions on the Fock states and show that edge modes appear at the ends of the system, when open boundary conditions are considered. Our result is an example of entangled many-body states in 1D systems of strongly interacting particles, without requiring the spin, long-range microscopic interactions or external fields. Instead, the entanglement can be tuned by the empty space in the system, which determines the spatial freedom of the interacting particles.
}

\begin{document}

\authorrunning{I.Kleftogiannis and Ilias Amanatidis} 
\titlerunning{Entanglement and edge effects in 1d Hubbard systems}
\maketitle

\section{Introduction}

Many-body systems often exhibit novel properties, due to
the collective behavior of their many interacting components
that self-organize in unique ways.
As it is usually quoted, a many-body
system is more than just a sum of its individual parts.
This extra ingredient due to collectiveness that is difficult to
extract reductively, can lead to extraordinary phenomena, such as, the fractionalization
of the elementary electron charge in the fractional-quantum-Hall
effect~\cite{Tsui,Laughlin}. Other celebrated examples where many-body interactions lead to measurable
consequences are for example, superconductivity and spin liquids\cite{Bardeen,Savary}.
During the last decades, it was realized that not all quantum phases of matter, due to many-body interactions,
can be described by the well known Landau symmetry-breaking theory of phase transitions,
which involves local order parameters characterizing different phases of matter.
For example, the superfluid phase of a spin liquid and the fractional-quantum-Hall phase, possess properties like long-range spatial correlations
related to quantum entanglement\cite{kitaev1,amico,horodecki,haldane0,AKLT,Kitaev4,Calabrese,Pollmann,Chen,kim2,kim1,wang2}.
This reveals that in certain cases, quantum many-body systems exhibit richer self-organization
phenomena, than their classical counterparts, due to the strong quantum correlations
present. For example, entanglement is a crucial mechanism, 
when examining topologically ordered phases\cite{amico,Gu,Kitaev2,Kitaev3,Levin,Li},
that occur in gapped quantum many-body systems, whose ground state is highly degenerate. Topological order leads to robust states that have been demonstrated to be useful in fault-tolerant quantum computation~\cite{Kitaev3}.
The realization that such quantum phases
of matter exist, which do not obey
the traditional Landau theory, opens a huge field for the investigation of other
types of entangled phases, not necessarily topological in nature.
In order to quantify the strength of the entanglement
in many-body systems, several measures can been used,
like the entanglement entropy and the entanglement spectrum~\cite{Kitaev2,Levin,Li,Hamma,Popkov,Alba}.
Some studies of entanglement and relevant topological features in Hubbard models have been studied in\cite{Farias,Ren,Ejima,Yoshida,Raghu,chung}.
Also several experimental realizations of quantum many-body systems with interesting topological and entanglement features, have been accomplished in cold-atomic and photonic systems\cite{yan,xu,xyguo,bernie,roushan}.

In this paper we investigate the many-body orders in the ground state of strongly interacting spinless particles confined in 1D.
We consider superpositions of many-body Fock states with spatial constraints, or equivalently,
collections of point-like(localized) particles arranged at different
spatial configurations.
The many-body orders of these states
are determined by the spatial freedom of the particles
according to the system filling.
These states can be realized in spinless Hubbard chains of hardcore bosons with strong short-range interactions. In the ground state of this system the particles arrange in 
Fock states that minimize the energy of the system, thus giving many degenerate
or nearly degenerate many-body ground states that are energetically
isolated from the other states of the system.
We calculate the spatial quantum correlations
that are created when the system is in a superposition of such many-body Fock states.
By splitting the system in two partitions we calculate analytically and
numerically the reduced density matrix and the bipartite entanglement,
which is a way to quantify the many-body correlations~\cite{amico,horodecki}.
Using this method, we show that the strength of entanglement varies
according to the spatial freedom of the particles,
determined by the system filling. In addition edge modes appear at the
ends of the system, when open boundary conditions are considered,
that can be correlated with each other due to the
entanglement. In overall, we demonstrate a simple mechanism
to create entangled states in the strong interaction or the charge density wave (CDW) limit of many-body systems, with controllable entanglement strength.

\section{Superpositions of Fock states with spatial constraints}
\begin{figure}
\begin{center}
\includegraphics[width=0.9\columnwidth,clip=true]{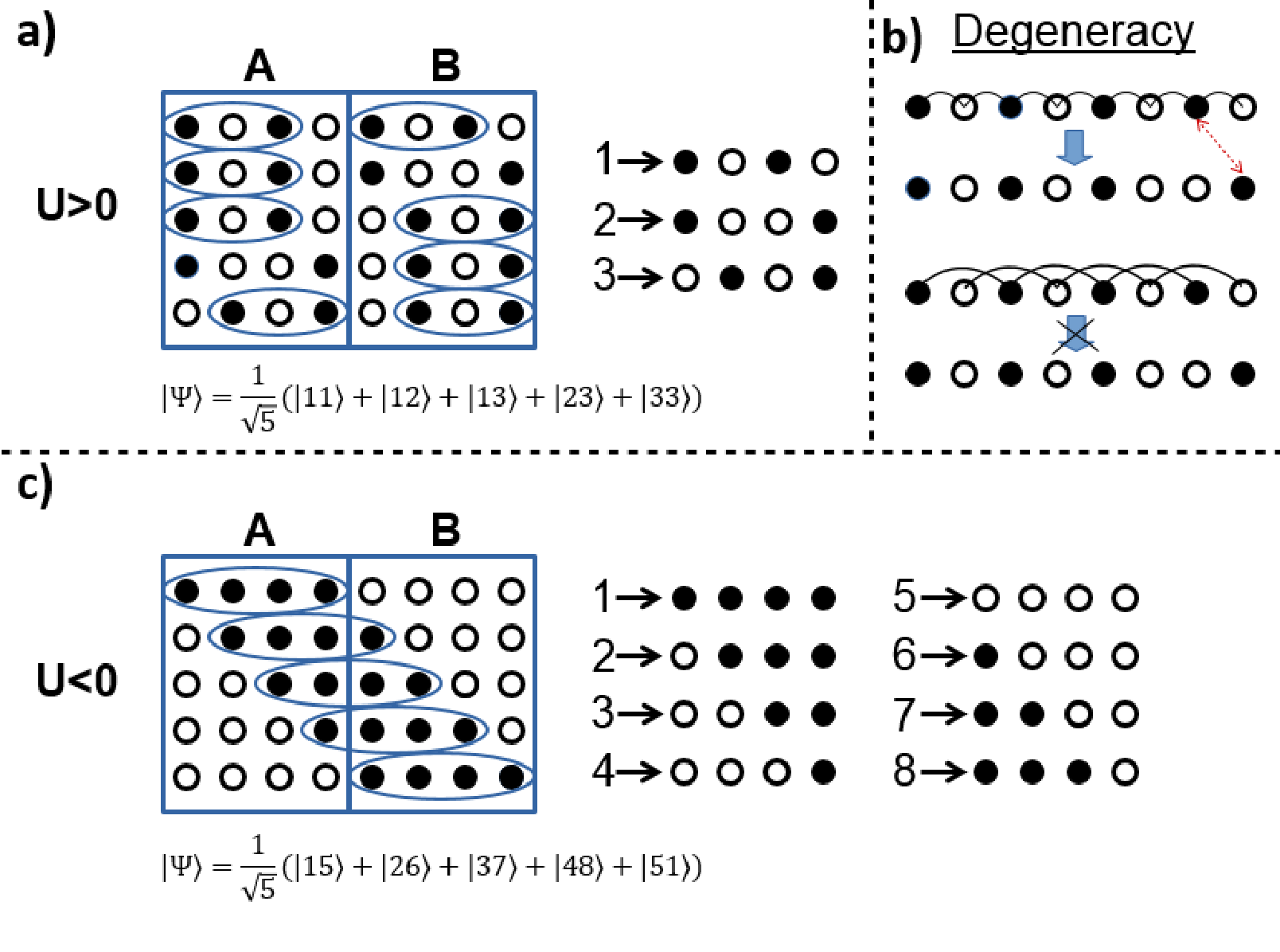}
\end{center}
\caption{a) The different particle configurations for the ground state of a half-filled
spinless Hubbard chain with nearest neighbor repulsive interaction $U>0$ and four particles (N=4).
Open(filled) circles denote unoccupied(occupied) sites. These microstates
minimize the system energy forming N+1 degenerate ground states.
The system can be split in two partitions A and B in order to calculate its
quantum correlations (entanglement).
b) First nearest neighbor hopping allows transition between the ground states
lifting their degeneracy, unlike second nearest neighbor hopping which preserves it.
c) The microstates for attractive interactions $U<0$ and $f=1/2$.
The stacking of all the particles minimizes their energy giving N+1 ground states.}
\label{fig_1}
\end{figure}

For our study, we consider collections of point-like(localized) particles arranged in different
spatial configurations.  A few examples
of such states can be seen in Fig. \ref{fig_1}(a).
These states are known also, as charge
density wave (CDW) states that can carry collectively charge current through the
system. The following spatial constraints in the self organization
of the particles are assumed.
Firstly, only one particle is allowed per site in an 1D chain
of empty sites, as in Hubbard models of hard-core bosons.
Following this rule there are many different ways to arrange the particles, according to the empty space in the system. If no interaction between the particles is assumed, then all the possible configurations
allowed will occupy the same energy E=0. In order
to split energetically these configurations, 
we can add an additional spatial constraint,
a nearest neighbor interaction between the particles. 
This can be expressed by a
Hubbard-like Hamiltonian term in a chain with M sites,
\begin{equation}
H_U = U\sum_{i=1}^{M-1} n_{i}n_{i+1}
\label{eq_1}
\end{equation}
where $n_{i}=c_{i}^{\dagger}c_{i}$ is the number operator, with $c_{i}^{\dagger},c_{i}$ being the  creation and annihilation operators for spin-less particles at site i.
The interaction strength U is measured in units of eV and can be either positive or negative for repulsive
or attractive interaction respectively.
The particles interact with strength U only when they are occupying neighboring sites
in the chain. We consider that the Hubbard chain terminates at sites 1 and M
with hard-wall/open boundary conditions.
The ratio of the number of particles $N$ over the number of sites in the Hubbard chain $M$,
determines the filling $\textit{f}=N/M$ of the system.
As we shall show the spatial freedom
of the particles, at different fillings, can lead to surprisingly complex
behaviors, even for this simple short-range interacting model.
We restrict our study on many-body wavefunctions that are symmetric under exchange of two particles. In addition we have assumed that the particles cannot occupy the same quantum state, i.e., only one particle is allowed per site (occupation number $n_{i}$ can be either 0 or 1).
This is the case of the hardcore bosons\cite{guo,wang1,zhang,Varney} that can be realized in cold atom and helium-4 systems experimentally~\cite{yan,islam,goldman,Bloch}.
The hardcore bosons satisfy the commutation relation $[c_{i},c_{j}^{\dagger}] = (1-2n_{i})\delta_{ij}$. In order to characterize the many-body states in the text
we use 1(0) to denote occupied(unoccupied) sites of the
Hubbard chain. Schematically we represent it with
filled(empty) circles.

In the rest of the paper we focus on the repulsive interaction case
$U>0$, although we present some brief analysis of the
attractive interaction case $U<0$ in section 5.
The eigenstates for $U>0$ and $\textit{f} \leq 1/2$  are determined by the possible Fock microstates that correspond
to the discrete energies $E=0,U,2U,...,(N-2)U,(N-1)U$.
Each energy contains states with different
microstructures.

In overall we can see that the interaction between the particles
splits the Hilbert space of the non-interacting system (U=0)
that contains all possible Fock states, in
subspaces containing states according to the different
particle configurations allowed at each energy.
For example the first excited states contain
at least one pair of particles at adjacent
sites, for instance the state $ | 10110100 \rangle$. This state has energy
$E=U$. In this paper we focus on the
ground states with energy $E=0$ where all particles
are seperated by at least one empty site, for example like the
state $ | 10101010 \rangle$ as shown in Fig. \ref{fig_1}(a).
These are many-body Fock states
where neighboring sites can never be simultaneously
occupied. This is valid when $\textit{f} \leq 1/2$ which concerns all the results we present in our paper.

The degree of degeneracy of the Fock states
depends on the filling $\textit{f}=N/M$,
which determines the spatial freedom of the interacting particles.
The number of allowed particle configurations/microstates can be expressed
mathematically with factorials via combinatorics, in the binomial form
\begin{equation}
D(N,M) = \binom{M-N+1}{N},
\label{eq_2}
\end{equation}
for the states that have at least one empty site between all the particles.

If there are $N$ particles in a system of $M=2N$ sites,
for half filling $\textit{f} = 1/2$,
then from Eq. (\ref{eq_2}) the number of degenerate ground states is N+1.
This can be seen in Fig. \ref{fig_1}(a) where we show a schematic representation
of the different states for a half-filled system
with $N=4$. Below half-filling $(\textit{f} < 1/2)$,
the particles are spatially less restricted and therefore the number of ground states
increases. As we shall show this degree of spatial freedom
affects the entanglement properties of the ground state.
We remark that such many-body states can be realized
experimentally in cold-atomic systems\cite{xu}.

The simplest way to construct the respective many-body wavefunction, is to assume
a linear superposition of the degenerate many-body states,
\begin{equation}
\ | \Psi_G \rangle =\frac{1}{\sqrt{D}} \sum_{i_{1}=1}^{M_1} \prod_{n=2}^{N}\sum_{i_{n}=i_{n-1}+1}^{M_n} [1-\delta_{i_{n},i_{n-1}+1}] | \{ i_1...i_n \} \rangle.
\label{eq_3}
\end{equation}
We have $M_{n}=M-N+n$ while index $i_n$
denotes the occupied sites in the Hubbard chain.
The list $\big\{ i_1,...i_n \big\}$ has length N
and determines a Fock state $| \{ i_1...i_n \} \rangle$.
This type of superposition with equal amplitudes for
each microstate is a reasonably good approximation
when the microstates are degenerate or nearly degenerate.
A similar wavefunction can be used for describing
the states of ferromagnetic spin chains with
periodic boundary conditions (PBC)\cite{Popkov}.
In addition Eq. (\ref{eq_3}) is a physically acceptable wavefunction
for the system described by  Eq. (\ref{eq_1})
since it results in spatially mirror symmetric particle density,
which can be defined as,
\begin{equation}
\ \overline{n_{i}} = \langle \Psi_G| n_{i} |\Psi_G \rangle.
\label{eq_4}
\end{equation}
The wavefunction  Eq. (\ref{eq_3}) satisfies
$\overline{n_{i}} = \overline{n_{M-i+1}}$.
Eq. (\ref{eq_3}) can be thought as an explicit ansatz for the ground state
wave function of Eq. (\ref{eq_1}).

Other statistical mixtures giving different amplitudes
are also possible instead of Eq. (\ref{eq_3}). For example, the amplitudes will be modified
when considering a hopping term of the type
\begin{equation}
H_t = t_{1} \sum_{i=1}^{M-1} c_{i}^{\dagger}c_{i+1} + t_{2} \sum_{i=1}^{M-2} c_{i}^{\dagger}c_{i+2}+ h.c.
\label{eq_5}
\end{equation}
in addition to the interaction term Eq. (\ref{eq_1}), where
$t_{1}$ is a nearest-neighbor hopping and $t_{2}$ is a next nearest-neighbor hopping. Note that in general there will be strong mixture between the Fock states at E=0 and the excited states, when adding this type of hopping term. However we expect the states studied in our paper to be
a good approximation for strong interaction strength $U  \gg t$.  

The effect of a weak first or second nearest
hopping on the ground states of Eq. (\ref{eq_1})
can be understood perturbatively.
Consider the half filled case.
First nearest neighbor hopping will allow transitions
between the Fock states, creating a small dispersion
that will lift their degeneracy.
For example, acting with a hopping term $c_{6}^{\dagger}c_{5}$
on the ground state $ | 10101010 \rangle$
results in the state $ | 10101001 \rangle$, as can be seen in Fig. \ref{fig_1}(b).
These degenerate ground states can be thought as the different
sites in a tight-binding chain.
A hopping between the sites will create
a dispersion in the energy spectrum of the chain, generating a band structure,
which is equivalent to lifting the degeneracy of the ground states for $t_1=0$.
On the other hand acting with a second nearest neighbor hopping
$c_{i}^{\dagger}c_{i+2}$ on the ground states,
for any i, will result only in excited states, which contain at least
a pair of particles occupying neighboring sites, contributing energy U,
the energy gap that separates them from the ground states.
Therefore the degeneracy of the ground states will not be lifted when adding weak second nearest neighbor hopping.
Also, the particle configurations for each ground state
will be preserved. Therefore, if the second
nearest neighbor $t_2$ is considered only, then the Fock states
remain degenerate at energy $E=0$ for half filling. This mechanism is demonstrated
schematically in Fig. \ref{fig_1}(b).

We have verified the above results by applying degenerate first order perturbation theory.
We have found that the energy of the perturbed system with first nearest neighbor hopping will be,
\begin{equation}
  E_{G}(j) \sim E_{G}^{0} + 2t_{1}cos\left(\frac{\pi j}{D+1}\right ),
  \label{eq_6}
\end{equation}
where D is the degeneracy (Eq. (\ref{eq_2})) and j is an integer taking values j=1,2..D,
for each of the perturbed ground states.
This is the energy dispersion of a tight-binding chain with D sites and hardwall boundary conditions.
Moreover each of these perturbed ground states can be written
as linear combination of the unperturbed ones. The amplitudes are
given by the corresponding wavefunction for a state j in the tight-binding chain,
\begin{equation}
  |\Psi_{G}(j)\rangle = \sqrt{\frac{2}{D+1}} \sum_{x=1}^{D}sin\left(\frac{\pi j x}{D+1}\right )|x\rangle,
  \label{eq_7}
\end{equation}
where $|x\rangle$ is the lattice site, representing each of the unperturbed states $| \{ i_1...i_n \} \rangle$ in the linear combination Eq. (\ref{eq_3}),
running over the degenerate space of dimension D.
Both results above are well known solutions for a XXZ spin
chain, to which our model can be mapped at half-filling\cite{spinhaldane}.

\begin{figure}
\begin{center}
\includegraphics[width=0.9\columnwidth,clip=true]{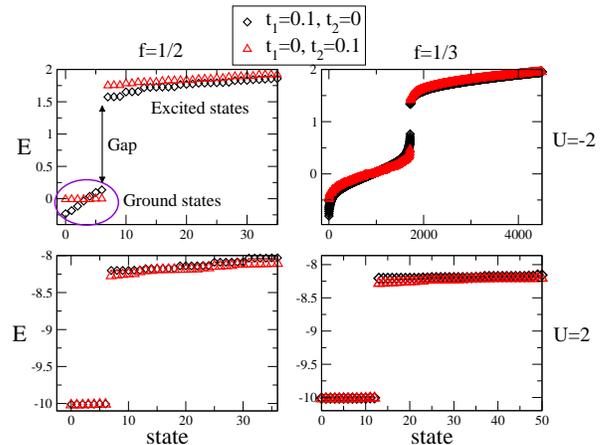}
\end{center}
\caption{The energy levels for repulsive ($U>0$) and attractive ($U<0$) interaction
with first ($t_1$) or second ($t_2$) nearest neighbor hopping.
In all cases the ground states are isolated
from the excited states since they are separated by a large energy gap.
For repulsive interaction the ground state degeneracy is preserved only for
$t_1=0,t_2=0.1$. }
\label{fig_2}
\end{figure}

In order to check the effect
of the hopping term on the ground state degeneracy, we have calculated the energy spectrum of the Hamiltonian
including both the interaction term Eq. (\ref{eq_1}) and the hopping term Eq. (\ref{eq_5}),
using numerical diagonalization, for a system with N=6 and $U=2$ and for different fillings.
The eigenvalues can be seen in Fig. \ref{fig_2}. As we have argued above, for half-filling ($\textit{f} =1/2$) the degeneracy is preserved for $t_1=0eV,t_2=0.1eV$
while it is lifted for $t_1=0.1eV,t_2=0eV$.
Although the degeneracy is lifted also for lower fillings ($\textit{f} =1/3$), the energy gap from the excited states does not close. Note that the hopping
we considered causes mixing of the ground states with the excited states unless $U>>t_1,t_2$.

In our analysis we consider a superposition of Fock states described by
Eq. (\ref{eq_3}), which are the ground states of Eq. (\ref{eq_1}).
We assume that an infinitesimally small hopping ($|U| \gg t$), for
example one that does not create clustering of the particles,
will not affect the micro-structures of these states, but might lift
their degeneracy, as we have shown above. As long as the gap from the excited states
at energy $E=U$ does not close, we can consider these nearly degenerate
ground states as an isolated Hilbert space and study its entanglement properties
by using the wavefunction Eq. (\ref{eq_3}) or Eq. (\ref{eq_7}).
Our approach can be thought as a good approximation to estimate the entanglement
properties, when spatial restrictions are imposed in the self-organization of the particles via different interaction terms.
We expect that for a sufficiently strong nearest neighbor interaction term, the major contribution in the ground state to come from Fock states where neighboring sites can never be simultaneously occupied like those in Eq. (\ref{eq_3}).

A few notes about the properties
of the system that we have considered. If we consider our system as a quantum fluid, then the encircled areas in Fig. \ref{fig_1}a would be incompressible inside each partition, since bringing two particles on neighboring
sites requires overcoming the energy gap U.
The half-filled system can be mapped onto a XXZ chain
of spins S=1/2\cite{spinhaldane}. By using this analogy we can see that the ground state of our model contains hidden anti-ferromagnetic order.
This can be understood easily by labeling occupied(unoccupied) sites as
spin up(down). Then the spin alternates between up and down, as in a chain containing anti-ferromagnetic
order. This is essentially a consequence of the
microscopic rule that there must be at least
one unoccupied site between all the particles.
The idea can be applied to any filling, since we can condense successive unoccupied sites into one.

\section{Entanglement}
The entanglement in the strong interaction or the CDW limit,
is governed by the different ways the particles organize to form the ground states.
In essence, the quantum superposition of the
enumerative combinatorics of the point-particles,
described by Eq. (\ref{eq_3}), is creating quantum mechanical
correlations in the system.
Therefore we can estimate the entanglement, without diagonalizing
the full Hamiltonian of the system,
but by examining the microstructures inside the Fock states that contribute in the ground state wavefunction.
The advantage of this approach compared to other methods like numerical diagonalization of the full Hamiltonian,
is that we can study a relatively larger number of particles.

A well established approach to quantify the quantum correlations is to split a system in two partitions, say A and B forming this way a composite system. Then
the entanglement between these partitions can be estimated via the reduced density matrix of partition A, $\rho_{A}\equiv tr_{B}|\Psi\rangle \langle\Psi|$
after tracing out the rest of the system, that is partition B. The elements of the reduced density matrix can be calculated via
$\rho_{A}^{ij}=\sum_{k \in B} \Psi_{ik}^{*} \Psi_{jk}$, where $\Psi_{ik}=\frac{1}{\sqrt{D}}$ is the amplitude
for each partitioned ground state $|ik \rangle$, where i(k) is the corresponding
microstate in A(B).
Moreover the von Neumann entanglement entropy can be calculated
\begin{equation}
S_{A}\equiv -tr(\rho_{A}ln(\rho_{A})).
\label{eq_8}
\end{equation}
The scaling of the entanglement entropy provides
information about the strength of entanglement in 1D quantum systems.
For critical 1D phases it has been shown that
the entropy diverges logarithmically with increasing
the partition size, while it saturates/converges
for non-critical phases\cite{kitaev1}. The logarithmic
divergence implies stronger entanglement than the
converging case.

In the case of half-filling, the density matrix
obtains a simple form that allows the calculation of
the entanglement entropy analytically as follows.
We start by noticing that each subsystem A or B is half the size of the
full composite system, with $N/2$ particles distributed
in $N$ sites. In order to calculate the element $\rho_{A}^{11}$
of the density matrix, we can fix the microstate inside A at $|1 \rangle$
and then count the different microstates in B.
These are $N/2+1$, as if B was an isolated system and we wanted to obtain the number of its
ground states. Then we multiply by the square modulus of the  normalization
factor $\frac{1}{\sqrt{D}}=\frac{1}{\sqrt{N+1}}$ and obtain $\rho_{A}^{11}=\frac{N +2}{2(N+1)}$.
All the other elements of the density matrix $\rho_{A}^{ij}$
for $i,j \neq 1$ are equal to $\frac{1}{N+1}$,
since the rest of the microstates in A appear
only once in the ground state of the composite system.
For convenience we define  $x_{1} =  \frac{N +2}{2(N+1)} $ , $x_{2} =  \frac{1}{N+1} $.
Then the only two non-zero eigenvalues of the reduced matrix $\rho_{A}$ are
\begin{equation}
\rho_{A_{1,2}} =  \frac{1}{2} \left( x_{1} + a x_{2} \pm  \sqrt{(x_{1} -ax_{2})^2 + 4ax_{2}^2 }  \right),
\end{equation}
where $a = \frac{N}{2} $.
The  entropy of the subsystem A in terms of the eigenvalues of the density matrix becomes
\begin{equation}
S_{A} =- \rho_{A_1}ln\rho_{A_1} - \rho_{A_2}ln\rho_{A_2}.
\label{eq_9}
\end{equation}
At the thermodynamic limit $N\rightarrow \infty$ the above eigenvalues
become $\rho_{A_{1,2}} =1/2$, resulting in the entropy
\begin{equation}
S_{A} = ln2.
\label{eq_10}
\end{equation}

\begin{figure}
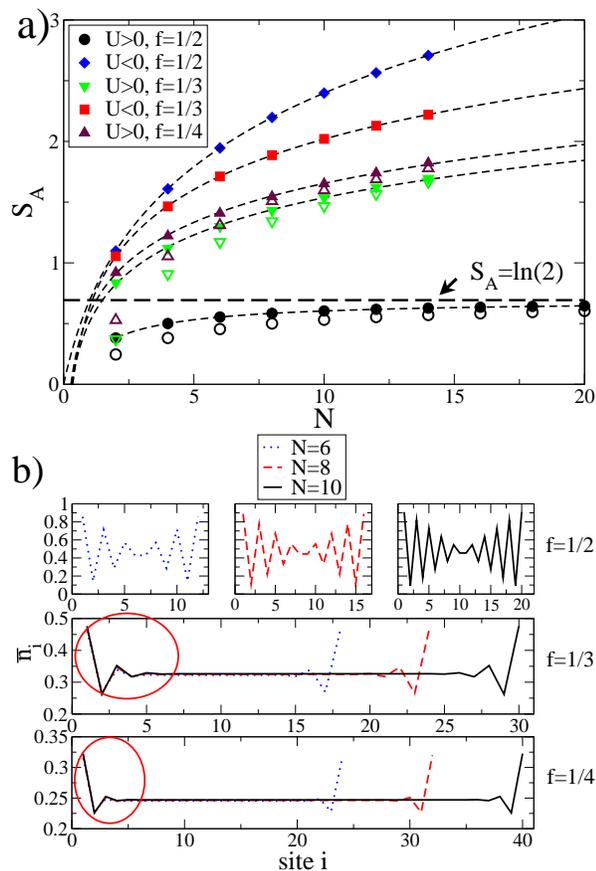

\begin{center}
\includegraphics[width=0.7\columnwidth,clip=true]{fig3a.eps}
\includegraphics[width=0.7\columnwidth,clip=true]{fig3b.eps}
\caption{a)The entanglement entropy for a partition A of the full system,
at different fillings $\textit{f}$ vesrus the number of particles N.
For half-filling $\textit{f}=\frac{1}{2}$ the entropy converges at $\ln{2}$
as two spins in a singlet maximally entangled Bell state. For lower fillings
the entropy diverges logarithmically as $S_A\sim\ln{N}$, implying stronger entanglement.
The last two points in the curve for  $U>0$ for $\textit{f}=1/4$ are calculated
via extrapolation. The filled(open) triangles and circles, are calculated
using Eq. (\ref{eq_3})(Eq. (\ref{eq_7})) b) The occupation probability $\overline{n_{i}}$ (particle density) versus each site i in the Hubbard chain, for N=6,8,10 and different fillings.
Edge modes can be seen for $\textit{f}<1/2$,
with uniform bulk density $\overline{n_{i}} \approx f$
and edge density that is a fraction $\overline{n_{edge}} \approx \frac{\textit{f}}{1-\textit{f}}$.
The edge modes are not affected by N.}
\label{fig_3}
\end{center}
\end{figure}

This is the entanglement entropy value for a maximally
entangled Bell state of two spins in the singlet state.
The convergence at the thermodynamic limit
implies semi-local correlations,
that result in weak entanglement as in the non-critical
phases of XY spin chains~\cite{kitaev1}.
This result agrees with the numerical calculation of
$S_A$ versus N using the ground state Eq. (\ref{eq_3}) , shown in Fig. \ref{fig_3}(a)
, where the dashed line is the analytical result Eq. (\ref{eq_6}).
Also, we have plotted $S_A$ using the amplitudes
of Eq. (\ref{eq_7}) for the lowest energy state with $j=D$,
shown as empty circles, which gives the same result as the $j=1$ case.
At the thermodynamic limit both cases, converge asymptotically to the 
value $S_{A} = ln2$.

Note that the steps above
are valid for even number of particles N,
half of which go at each partition.
Following a similar method
we can derive that $S_{A} = ln2$
for odd N also.

The origins of the weak entanglement can be understood as follows.
We could imagine the partitions A and B,
as two isolated many-body systems. Each one is half the size of the full composite system, with $N/2$ particles distributed among $N$ sites.
As for the full system, the ground states of each partition
are determined by the different microstates that
minimize the energy, which requires that
there must be at least one empty site between two particles.
These states form the Hilbert space $\mathcal{H}^{A(B)}_{G}$
of each isolated partition. The short range interaction plays an important role at the boundary between A and B,
when they are put together to form the full system.
As an example consider the system shown in Fig \ref{fig_1}(a). Each partition
contains two particles and four sites. When we form the ground state
of the full system, the combinations of states $ | 1001 \rangle | 1010 \rangle$,
$ | 1001 \rangle | 1001 \rangle$, $ | 0101 \rangle | 1010 \rangle$ and
$ | 0101 \rangle | 1001 \rangle$ have to be excluded.
In other words, the ground state Hilbert space of
the full system $\mathcal{H_{G}}$ is not simply the tensor product
of the individual spaces of A and B, that is, $\mathcal{H_{G}} \neq \mathcal{H}^{A}_{G} \otimes \mathcal{H}^{B}_{G}$.
The two partitions become weakly entangled, due to the local particle interaction at their boundary.

In Fig. \ref{fig_3}(a) we show $S_A$ for lower fillings ($\textit{f} =1/3,1/4$) calculated
numerically using the ground state Eq. (\ref{eq_3}).
For all these cases $S_A$ diverges logarithmically,
as in the critical phases of XY spin chains~\cite{kitaev1,Calabrese},
implying stronger entanglement than $\textit{f} =1/2$.
The same logarithmic behavior is observed if we use Eq. (\ref{eq_7})
to calculate $S_A$ (for $j=1$ and $j=D$), represented by empty triangles in Fig. \ref{fig_3}(a).
This is an indication of the larger complexity of the
particle configurations due to the larger spatial freedom
compared to the half-filling. We remark that for $\textit{f} <1/2$,
the particle number at each partition
is not conserved. Therefore we cannot form the ground state Hilbert space of
the full system $\mathcal{H_{G}}$ as a tensor product
of the individual spaces of two isolated partitions A and B, even if we remove the interaction
at the boundary between them, unlike the half-filled case.
This means that we cannot completely remove the entanglement
with local operations, which is another indication of the strong
entanglement, that induces long-range spatial correlations.

The different entanglements are related to
the spatial freedom of the particles
at the respective fillings.
In a low filled system we can freely add an additional
particle without requiring more energy, as long as there is
at least one unoccupied site between it and the
rest of the particles. This way we can fill the empty space
of the system with $Int[(M+1)/2-N]$ particles (for even M)
that will go at the ground state without affecting
its energy, or the energy gap from the excited states.
In this sense, the low fillings
in our model correspond to a transition towards
a superfluid phase\cite{hen}, i.e. the system at low fillings lies in a critical regime.
This could explain the logarithmic divergence of the
entanglement entropy with the partition size,
which occurs in general at the critical regime of 1D many-body
systems. In the half-filled case on the other hand
the system lies in a Mott insulating phase, since we cannot add an additional particle without exciting the system.
Therefore the different scaling behavior
of the entanglement entropy for the half-filled
and low filled cases, is related to the different
phases the system lies at these fillings.

Note that a logarithmic divergence of the entropy is also
expected when U=0, i.e., by removing the spatial constraint
of an empty site between all the particles, leaving
only the hardcore boson constraint of one particle per site. In this
case all possible spatial configurations of the particles are allowed in Eq. (\ref{eq_3})
, which will contain the full entanglement of the whole system.
Our approach allows additional control of the entanglement,
due to splitting of the system in isolated Hilbert spaces that contain different entanglement strengths, according to the microstructure inside the corresponding Fock states.

Another way to detect the entanglement is by measuring the purity
of the quantum state with density matrix $\rho$ which can be quantified by $Tr(\rho^{2})$\cite{islam}.
When $Tr(\rho^{2})<1$ the state is mixed which means that it is not quantum
mechanically fully consistent and contains statistical fluctuations.
When this happens for the reduced density matrix
of a partition of a quantum system, entanglement with the rest of the system is implied.
We have found $Tr(\rho_{A}^{2})<1$ for all the fillings studied ($\textit{f} =1/2,1/3,1/4$),
which is an additional indication of the entanglement present in our system.

Our results bear resemblance to topological ordered phases of matter,
which is usually identified through its highly degenerate
ground state and strong long-range entanglement properties.
However we are not able to obtain both these properties simultaneously.
For example the half-filled system has a highly degenerate
ground state, that is not lifted by a second
nearest neighbor hopping, but lacks the strong entanglement.
On the other hand at lower filling the degeneracy is lifted for
nearest and second nearest hopping,
but the entanglement of these ground states, becomes stronger,
as indicated by the logarithmic divergence of the entanglement entropy.

In addition we have found that most of the eigenvalues of
the density matrix are doubly degenerate
as in the Haldane phase of spin chains where string order
and entanglement are present\cite{Calabrese,Pollmann,Li}.

\section{Edge effects}
In order to obtain additional features of the ground states we
calculate the occupation probability (particle density) for each site
of the Hubbard chain, using Eq. (\ref{eq_4}).

In Fig. \ref{fig_3}(b) we show $\overline{n_{i}}$ for different fillings
and number of particles N=6,8,10. For the half-filled case we observe fluctuations of the probability, as expected\cite{Nishimoto} due to the edges acting as impurities causing Friedel oscillations, that are larger at the two ends of the Hubbard chain. For the lowest fillings ($\textit{f} = 1/3,1/4$) however, the fluctuations smoothen out in the bulk of the Hubbard chain resulting in uniform density $\overline{n_{i}} \approx f$.

In order to understand this uniform bulk density
we can avoid the edge effects by applying periodic boundary conditions(PBC)
in the system. This can be done by changing the
upper summation limit to M in Eq. (\ref{eq_1}) and considering the condition $M+1 \rightarrow 1$.
Since we are interested in the $E=0$ states, we must remove the states that contain particles at the two edge sites of the chain  $ | 1010...01 \rangle$ as these will lift
the energy of the system by $E=U$. We can use Eq. (\ref{eq_2}) to calculate the number
of these states, which is $D(N-2,M-4)$. Therefore
the number of ground states with PBC is $D_{PBC}(N,M)=D(N,M)-D(N-2,M-4)$.
The corresponding particle density can be calculated by noticing that
for $\textit{f} \leq 1/2$ every occupied site has to lie
between two empty sites in a state of the type $ |...010...\rangle$.
Therefore, the density on every site, at the thermodynamic limit $N\rightarrow \infty$, becomes $\overline{n_{bulk}}=\frac{D(N-1,M-3)}{D_{PBC}(N,M)}=f$,
as shown in Fig. \ref{fig_3}(b) for low fillings.
However when edges are present
edge modes are formed due to the inclusion
of the states of the type $ | 1010...01 \rangle$, that
have a particle on both edge sites.
In this case, the occupation probability at the edge
can be obtained by the ratio between the states
that have one occupied site at the corresponding edge (site 1 or M) $D(N-1,M-2)$,
over the total number of microstates D(N,M), giving the edge density $\overline{n_{edge}}=\frac{D(N-1,M-2)}{D(N,M)}$.
Using Eq. (\ref{eq_2}) we find that $\overline{n_{edge}}=\frac{N}{1+M-N}=\frac{1}{\frac{1}{N}+\frac{1-\textit{f}}{\textit{f}}}$
which becomes a fraction $\overline{n_{edge}}=\frac{\textit{f}}{1-\textit{f}}$ in the thermodynamic limit $N\rightarrow \infty$. This result agrees well with the
edge density shown in Fig. \ref{fig_3}(b), even for the relatively
small number of particles shown (N=6,8,10).
We note that the dip of $\overline{n_{i}}$ near
the edge sites is due to the repulsive interaction U which
reduces the probability to find a particle on the neighboring site to the edge, if the edge site is already occupied.

The edge modes could be considered as excitations
of the particle density which remains uniform at the bulk
of the Hubbard chain. Each edge is contained
in one of the partitions A or B which are quantum mechanically correlated/entangled, as we have shown. Therefore, we can assume that the edge modes at the opposite ends of the system can be entangled with each other.

Combining the edge modes with the results
obtained in the previous section, namely the
ground state degeneracy and the variable entanglement,
it is tempting to assume that our system contains
topological order. However further analysis
is needed, through the calculation of topological
non-local measures like the string-order parameter 
for example\cite{Calabrese,Pollmann,Li}.

\section{Attractive interactions}
We briefly analyze the case of negative U, that is, for
attractive interactions between the particles. In this case
the ground state is obtained by stacking all the particles together
at neighboring sites, which minimizes their energy at $E=-U(N-1)$.
These states are separated by a large gap U from the first
excited states. An example of the microstates can be seen in Fig. \ref{fig_1}(c).
The degeneracy of these ground states is preserved for both
first and nearest neighbor hopping as can be seen in Fig.  \ref{fig_2},
unlike the repulsive interaction case. For the half-filled
case the entanglement entropy in Fig. \ref{fig_3}
follows the limit where the number of the different particle configurations in subsystems A and B
is equal to the corresponding number of microstates. At this limit the reduced density
matrix $\rho_{A}$ is diagonal, resulting
in maximum entropy $S_A=ln(N+1)$ which is the corresponding dashed curve in Fig. \ref{fig_3}(a).
For lower fillings the entropy is reduced
but still diverges logarithmically.

\section{Summary and Conclusions}
To summarize, we have investigated the self-organization of
strongly interacting spinless particles confined
in one-dimension. The spatial freedom of the particles
allows them to organize in various spatial configurations
(Fock states), corresponding
to different energy bands, that are separated by large gaps.
These states can be realized in spinless Hubbard chains
of hardcore bosons, which have gapped energy spectrum
with a highly degenerate ground state.

By considering a superposition of the many-body Fock states
for the lowest energy of the system (ground state),
we have calculated the resulting quantum correlations.
We have done that by splitting the system in two partitions
and then calculating the scaling of the entanglement entropy
of each partition with the system size.
We have shown that the strength of the entanglement
depends on the spatial freedom of the particles,
which is determined by the system filling.
For a half-filled system,
the entanglement resembles that of two spins
in a singlet maximally entangled Bell state.
At low fillings the entanglement
becomes stronger as indicated
by the logarithmic scaling of the entanglement entropy
of each partition. In addition we found that confining 
the system by using open boundary conditions, leads to
edge modes at the ends of the system at low fillings. These edge modes
can be correlated with each other,
due to the strong long-range entanglement appearing at 
the respective fillings.

Our results show that 1D Hubbard models of spinless hardcore bosons,
at the strong interaction limit or the charge density wave limit
can have spectrally isolated ground states with
entanglement determined by their filling,
despite the strong localization of
the particles. These orders are created due to the
spatial freedom of the particles which result in
formation of many-body states with rich micro-structures.
Essentially this mechanism
could be used as a way to tune the entanglement in this
type of systems, by controlling the empty space in the system,
without the application of external fields.

A natural extension of the current analysis would be to
investigate the many-body orders and the entanglement
in the different excited states that are separated by energy gaps,
which contain many-body states with richer structures. In addition
it would be interesting to examine the relevant many-body orders
in systems of higher dimension.

As a final note, we would like to remark that
our results contribute to the idea, that entangled states
with non-trivial features, can be created via the collective behaviors
in many-body systems with relatively simple microscopic rules.
Such well known examples are for instance, the Haldane phase of integer spin chains\cite{haldane0,AKLT},
the Majorana modes in the Kitaev chain\cite{Kitaev4} and the topological order in the toric code\cite{Kitaev3}.
Apart from its fundamental significance, this approach might be useful in the on-going research
on the experimental realization of different topological phases
in cold-atomic and photonic systems.

\begin{acknowledgement}
We would like to  thank Ara Go, Guang-Yu Guo, Alexander Schnell, Vladislav Popkov
and Daw-Wei Wang for useful comments. We acknowledge resources and financial support provided by
the National Taiwan University, the Ministry of Science and Technology,
the National Center for Theoretical Sciences of R.O.C. Taiwan and the
Center for Theoretical Physics of Complex Systems in Daejeon Korea under the project
IBS-R024-D1.
\end{acknowledgement}

%
%

\end{document}